\documentclass[aps,showpacs,amsmath,amssymb,prl,twocolumn]{revtex4}
\usepackage{pslatex}
\usepackage{bm}
\usepackage{graphicx}

\begin{document}
\title{Weak compressible magnetohydrodynamic turbulence in the solar corona}
\author{Benjamin D. G. Chandran}
\affiliation{Space Science Center, Institute for the Study of Earth, Oceans, and Space,
University of New Hampshire, Durham, NH}
\begin{abstract}
This Letter presents a calculation of the power spectra of weakly
turbulent Alfv\'en waves and fast magnetosonic waves (``fast waves'')
in low-$\beta$ plasmas. It is shown that three-wave interactions
transfer energy to high-frequency fast waves and, to a lesser extent,
high-frequency Alfv\'en waves. MHD turbulence is thus a promising
mechanism for producing the high-frequency waves needed to explain the
anisotropic heating of minor ions in the solar~corona.
\end{abstract}
\maketitle

The heating of the solar corona is a long-standing 
problem.  Measurements taken with the Ultraviolet Coronagraph
Spectrometer (UVCS) have provided important constraints on coronal
heating, showing, for example, that $T_\perp \gg T_\parallel$ and
$T_\perp \sim 10^8$~K for O$^{+5}$ ions at a heliocentric distance~$r$
of roughly two solar radii, where $T_\perp$ and $T_\parallel$ are the
temperatures for random motions perpendicular and parallel to the
magnetic field~$\bm{B}$.~\cite{kohl98,ant00} These measurements imply
that the average magnetic moment $k_BT_\perp/B$ of O$^{+5}$ ions
increases rapidly with~$r$ and show that O$^{+5}$ ions are heated by
electric and magnetic-field fluctuations with frequencies~$\omega$
comparable to or greater than the ions' cyclotron
frequency~$\Omega$. (If $\omega \ll \Omega$, the average magnetic
moment is almost exactly conserved.)

Different sources have been proposed for the required high-frequency
waves, including reconnection events in the coronal
base~\cite{hol02,axf92,cra00,tum01}, heat-flux-driven plasma
instabilities~\cite{mar04}, and magnetohydrodynamic (MHD)
turbulence~\cite{col68,dmi02,cra05}. An apparent difficulty with this
last source is the finding that in incompressible and weakly
compressible MHD turbulence there is little or no cascade of energy to
high frequencies.~\cite{she83,gol97,ng97,gal00,lit03} However,
incompressible and weakly compressible MHD neglect the fast
magnetosonic wave (``fast wave''). In this Letter, a weak turbulence
calculation is used to show that when fast waves are accounted for,
MHD turbulence in low-$\beta$ plasmas transfers energy to
high-frequency fast waves and, to a lesser extent, high-frequency
Alfv\'en waves.  (In the corona, $\beta \equiv 8\pi p/B^2\sim 0.01$,
where~$p$ is the pressure.) The high-frequency waves produced by MHD
turbulence may be of importance not only for coronal heating, but for
particle acceleration in solar flares as well.~\cite{mil96}

The MHD momentum and induction equations with Laplacian viscosity and
resistivity are
\begin{equation} 
\rho\left(\frac{\partial \bm{v} }{\partial t} + \bm{v} \cdot \bm{\nabla}
\bm{v} \right)  =  - \bm{\nabla} \left(p +\frac{B^2}{8\pi}\right) + \frac{\bm{B} \cdot
\bm{\nabla}\bm{B}}{4\pi} + \rho \nu \nabla^2 \bm{v} \label{eq:momentum} 
\end{equation} 
and
\begin{equation} 
 \frac{\partial \bm{B} }{\partial t} = \bm{\nabla} \times (\bm{v} \times \bm{B} )
+ \eta \nabla^2 \bm{B} ,
\label{eq:ind} 
\end{equation} 
where~$\rho$ is the density, $\bm{v}$ is the velocity, $p$ is the pressure, and~$\bm{B}$ is
the magnetic field. In this Letter, the magnetic field is taken to consist of a uniform
background field and a small-amplitude fluctuating field: $\bm{B} =
B_0 \hat{z} + \delta \bm{B}$.  The pressure is discarded
since $\beta$ is taken to be~$\ll 1$.
The spatial Fourier transforms of $\bm{v} $ and $\bm{b} \equiv \delta \bm{B}
/\sqrt{4\pi \rho}$ can be written
\begin{equation} 
\bm{v} _k = v_{a,k} \,\hat{\bm{e}}_{a,k} + v_{f,k}\,
\hat{\bm{k}}_\perp + v_{s,k}\hat{\bm{z}}
 \label{eq:pol1} 
\end{equation}
and
\begin{equation}
\bm{b}_k   =  b_{a,k} \,\hat{\bm{e}}_{a,k} + b_{f,k} \,\hat{\bm{e}}_{f,k},
\label{eq:pol2} 
\end{equation} 
where $\hat{\bm{e}}_{a,k} = \hat{\bm{z}} \times \hat{\bm{k}}_\perp$ is the
Alfv\'en-wave polarization vector at wave vector~$\bm{k}$,
$\hat{\bm{k}}_\perp = \bm{k}_\perp/k_\perp$, 
$\bm{k}_\perp = \bm{k} - k_z \hat{\bm{z}}$,
and $\hat{\bm{e}}_{f,k} = \hat{\bm{e}}_{a,k} \times \bm{k}/k$.
The Alfv\'en-wave amplitude is given by 
$a^\pm_k  =  v_{a,k} \pm b_{a,k}$,
and (since $\beta\ll 1$) the fast-wave amplitude is given by
$ f_k^\pm  =  v_{f,k} \pm b_{f,k}$.
Upon neglecting nonlinear terms in the momentum and induction equations, one obtains
$\partial a^\pm_k/\partial t  =  \pm ik_z v_A a^\pm _k$ and
$\partial f^\pm_k/\partial t   =  \pm ik v_A f^\pm _k$,
where $v_A= B_0/\sqrt{4\pi\rho}$ is the Alfv\'en speed.  Alfv\'en
waves have frequency~$\mp k_z v_A$ and propagate along magnetic
field lines. Fast waves have frequency~$\mp k v_A$ and can
propagate in any direction. The $v_{s,k}\hat{\bm{z}}$ term in
equation~(\ref{eq:pol1}) corresponds to the slow magnetosonic wave, which
has a frequency that approaches~zero as~$\beta\rightarrow 0$.

Weak turbulence consists of waves whose amplitudes are sufficiently
small that nonlinear interactions between waves can be treated as a
small perturbation to a wave's linear behavior. Weak turbulence theory
is based on the assumptions of random wave phases and approximately
Gaussian statistics.~\cite{zak92} These assumptions are problematic
for acoustic turbulence, because sound waves propagating
non-dispersively in the same direction interact coherently for long
times.~\cite{zak70,zak92} The same issue arises for fast waves,
because of their acoustic-like dispersion relation.  However,
fast-wave interactions with Alfv\'en waves and slow magnetosonic waves
limit the interaction time for pure fast-wave interactions, which may
allow weak turbulence theory to apply to MHD at low~$\beta$ even if it
does not apply to acoustic turbulence.  Although this issue remains
unresolved, weak turbulence theory is a valuable starting point for
this difficult problem.

To simplify the analysis, the slow magnetosonic wave is neglected, the
density~$\rho$ is taken to be a constant, and (to maintain energy
conservation when~$\rho$ is held constant) the $\bm{v} \cdot \nabla
\bm{v} $ term in equation~(\ref{eq:momentum}) is replaced with $\bm{
  v}^A \cdot \nabla \bm{v} $, where~$\bm{v}^A$ is the part of the
velocity associated with Alfv\'en waves.  A different approach was
taken by~\cite{kuz01}, who included slow waves but neglected
three-wave interactions that did not involve slow waves.  Further work
including all the nonlinearities is needed.  The Alfv\'en-wave and
fast-wave power spectra for homogeneous turbulence are defined through
the equations $\langle a^\pm_k (a^\pm_{k1})^\star \rangle = A_k^\pm
\delta(\bm{k} - \bm{k}_1)$, and $ \langle f^\pm_k (f^\pm_{k1})^\star
\rangle = F_k^\pm \delta(\bm{k} - \bm{k}_1)$, where $\langle \dots
\rangle$ denotes an ensemble average. It is
assumed that $A^+_k = A^-_k \equiv A_k$, that $F^+_k = F^-_k \equiv
F_k$, and that $\langle a^\pm_k f^\pm_{k1} \rangle = \langle a^\pm_k
f^\mp_{k1} \rangle = 0$.  Rotational symmetry about the $z$~axis is
also assumed, so that~$A_k = A (k_\perp, k_z,t)$ and $F_k = F(k_\perp,
k_z,t)$.  Taking the small-$\nu$ and small-$\eta$ limits and employing
the standard weak-turbulence approximations, one obtains the wave
kinetic equations,
\begin{widetext}
\begin{eqnarray} 
\hspace{-0.3cm} 
\frac{\partial A_k}{\partial t} 
 & = & \frac{\pi}{8 v_A } \int
d^3 p\, d^3q\, \delta (\bm{k} - \bm{p} - \bm{q}) \left\{ \delta
( q_z) 8(p_\perp n \overline{l})^2 A_q (A_p - A_k)
\hspace{0.1cm} +\hspace{0.1cm} \delta (k_z + p_z + q) M_1 [M_2 F_q(A_p
- A_k) + M_3 A_p(F_q - A_k)] \right. \nonumber\\ 
& & \hspace{0.33cm}
\left. + \hspace{0.1cm} \delta (k_z + p_z - q) M_4 [M_5 F_q(A_p - A_k) + M_3 A_p (F_q
- A_k)] \hspace{0.1cm} + \hspace{0.1cm} \delta (k_z + p - q) \;M_6 [M_7
F_q(F_p-A_k) + M_8 F_p(F_q - A_k)]\right\} \nonumber\\
& &\label{eq:wkA}\\
\hspace{-0.3cm} \frac{\partial F_k}{\partial t}  & = &
\frac{\pi}{8 v_A } \int d^3 p\, d^3q\, \delta (\bm{k} - \bm{p} - \bm{q}) 
\left\{
9\sin^2\theta[\delta (k-p-q)kqF_p(
F_q - F_k) \hspace{0.1cm} + \hspace{0.1cm} \delta (k+p-q)(k^2 F_p F_q +
kpF_q F_k - kqF_p F_k)] \right. \nonumber\\
& & \hspace{0.cm} 
+ \hspace{0.1cm} 
\delta(k-p_z + q_z) M_{9}[M_{10} A_q(A_p - F_k) + M_{11} A_p(A_q - F_k)]
\hspace{0.1cm}  + \hspace{0.1cm} 
\delta(k-p_z - q) M_{12}[M_{13}F_q(A_p - F_k) + M_{14}A_p(F_q - F_k)]
\nonumber\\
& & \hspace{0.cm} 
+ \hspace{0.1cm}
\delta(k+p_z - q) M_{15}[M_{16}F_q(A_p - F_k) +
M_{17}A_p(F_q - F_k)]  \left.\right\}, \hspace{0.1cm}
\label{eq:wkF} 
\end{eqnarray} 
\end{widetext}
where
\[
\hspace{-0.2cm}
\begin{array}{lcl}
M_2 &  = & -p_\perp m - (\cos\alpha + \,1/2)(k_\perp l +
p_\perp m + q_\perp n), \nonumber\\
M_3 & = & 2k_\perp l + 2p_\perp m + q_\perp n,\nonumber\\
M_5 & = & -p_\perp m + (\cos\alpha - \,1/2 )
(k_\perp l + p_\perp m + q_\perp n), \nonumber\\
M_7 & = & k_\perp \overline{l} (\cos\alpha - \,1/2) + p_\perp 
\overline{m}/2 \, + \sin\alpha \,\overline{n} (2p - \,q/2),\nonumber\\
M_{8} & = & k_\perp \overline{l} (\cos\psi +\, 1/2) -
\sin\psi \,\overline{m} (2q-\, p/2) -
q_\perp \overline{n}/2,\nonumber\\
M_{10} & = & p_\perp m + (\cos\theta + \,1/2)
(k_\perp l + p_\perp m + q_\perp n),\nonumber\\
M_{11} & = & q_\perp n + (-\cos\theta +\,1/2)
(k_\perp l + p_\perp m + q_\perp n),\nonumber\\
M_{13} & = & \sin\theta\, \overline{l} (-k + 2q) + p_\perp \overline{m} (\cos\theta
- \cos\alpha) + \sin\alpha\, \overline{n} (q - 2k),\nonumber\\
M_{14} & = & \sin\theta \,\overline{l} (k/2 - 2q) +
p_\perp \overline{m}
(- \cos\theta +\,1/2) - q_\perp \overline{n}/2,\nonumber\\
M_{16} & = & \sin\theta \,\overline{l} (-k + 2q) + p_\perp \overline{m} (\cos\alpha
- \cos\theta) + \sin \alpha \,\overline{n} (q-2k),\nonumber\\
M_{17} & = & \sin\theta \,\overline{l} (k/2\, - 2q) +
p_\perp \overline{m} (\cos\theta + \,1/2) -
q_\perp \overline{n}/2,\nonumber
\end{array}
\]
$M_1  =  M_2 + M_3$,
$M_4 =  M_5+M_3$,
$M_6  = M_7+M_8$, 
$ M_{9} =  M_{10} + M_{11}$,
$M_{12} =  M_{13}+M_{14}$, and
$M_{15} =  M_{16}+M_{17}$. The quantities
 $\theta$, $\psi$,
and~$\alpha$ are the angles between~$\hat{\bm{z}}$ and the wave
vectors~$\bm{k}$, $\bm{p}$, and~$\bm{q}$, respectively.  In the
triangle with sides of lengths $k_\perp$, $p_\perp$, and~$q_\perp$,
the interior angles opposite the sides of length~$k_\perp$, $p_\perp$,
and~$q_\perp$ are denoted $\sigma_k$, $\sigma_p$, and $\sigma_q$, and
$l=\cos\sigma_k$, $m = \cos\sigma_p$, $n=\cos\sigma_q$, $\overline{l}
= \sin\sigma_k$, $\overline{m} = \sin \sigma_p$, and $\overline{n} =
\sin\sigma_q$. The above form of the wave kinetic equation makes use
of the identities $k_\perp \cos(\sigma_q - \sigma_p) = q_\perp n +
p_\perp m$ and $k_\perp \sin(\sigma_q - \sigma_p) = q_\perp
\overline{n} - p_\perp \overline{m}$.

The right-hand sides of equations~(\ref{eq:wkA}) and (\ref{eq:wkF})
(the ``collision integrals'') represent the effects of resonant
three-wave interactions. The integrals sum over all possible
wavenumber triads, while the delta functions restrict the sum to
triads satisfying the resonance conditions $\bm{k} = \bm{p}+\bm{q}$
and $\omega_k = \omega_p + \omega_q$, where~$\omega_k$ is the
frequency at wavenumber~$\bm{k}$.  The equations~$M_1 = M_2+M_3$, $M_4
= M_5 + M_3$,~etc imply that $\partial A_k/\partial t$ (or $\partial
F_k/\partial t$) is positive at any wave number at which~$A_k$ (or
$F_k$) vanishes, provided the spectra are positive at other wave
numbers. The wave kinetic equations thus ensure that the spectra
remain positive (realizability).  Since the dissipative terms have not
been included, equations~(\ref{eq:wkA}) and (\ref{eq:wkF}) conserve
the energy per unit mass $E = \int d^3k\,(A_k + F_k)/2$ and have an
equipartition solution~$F_k= A_k=\mbox{constant}$.

The $\delta(q_z)$ in the collision integral of equation~(\ref{eq:wkA})
is equivalent to $2v_A \delta (k_zv_A - p_zv_A + q_zv_A)$ and represents
the frequency-matching condition
for resonant interactions involving three Alfv\'en waves
(``AAA interactions'').  The part of the collision integral that
contains this $\delta(q_z)$ is the same as the collision integral for
AAA interactions in incompressible MHD.  This term represents
interactions between oppositely directed Alfv\'en waves, in which the
field-line displacements caused by Alfv\'en waves travelling in one
direction along the magnetic field [represented by $A(q_\perp,
  q_z=0)$] distort Alfv\'en wave packets travelling in the opposite
direction. At $k_z=0$, only the AAA terms contribute to the right-hand
side of equation~(\ref{eq:wkA}), and the steady-state solution
$A(k_\perp, k_z=0) \propto k_\perp^{-3}$ can be obtained analytically
with the use of a Zakharov transformation, as in the
incompressible case~\cite{gal00}.  When $A(k_\perp , k_z=0)\propto
k_\perp^{-3}$, and when non-AAA interactions are neglected, a Zakharov
transformation yields $A_k \propto k_\perp^{-3}$ for any~$k_z$.  It
can be seen from equation~(\ref{eq:wkA}) that the time scale~$\tau_A$
for AAA interactions to transfer Alfv\'en-wave energy from $k_\perp$
to $2k_\perp$ is determined by~$A(k_\perp, k_z=0)$ and is independent
of~$k_z$, consistent with physical descriptions of the Alfv\'en-wave
cascade.~\cite{ng97,gol97,cha04} If 
the Alfv\'en-wave energy per unit mass~$(\delta v_{\rm rms})^2$ is
dominated by wavenumbers of order some characteristic wave number~$k_0$, if the spectrum is
quasi-isotropic at~$k\sim k_0$, and if $A(k_\perp, k_z=0)\propto
k_\perp^{-3}$ for~$k_\perp \gtrsim k_0$, then $\tau_A \simeq v_A/[k_\perp
(\delta v_{\rm rms})^2]$ for $k_\perp \gg k_0$, as in the
incompressible case.~\cite{ng97,gol97}

The terms in the collision integral of equation~(\ref{eq:wkF})
proportional to $\delta(k-p-q)$ and~$\delta(k+p-q)$ represent resonant
three-wave interactions involving only fast waves (``FFF
interactions'').  As can be seen from the delta functions, FFF
interactions occur only when $\bm{k}$ is parallel or anti-parallel to
both $\bm{p}$ and~$\bm{q}$, indicating that these interactions
transfer energy radially in $k$-space.  The FFF terms are the same as
the collision integral for weak acoustic turbulence~\cite{zak70}, up
to an overall multiplicative factor proportional to~$\sin ^2 \theta$,
and represent a weak form of wave steepening.  As $\sin\theta
\rightarrow 0$, the acoustic-like FFF interactions become less
efficient because the fast waves become less compressive.  If the
non-FFF terms are neglected, then a Zakharov transformation can be
used to show that $F_k = c_1 g(\theta) k^{-7/2}$ is a steady-state
solution to equation~(\ref{eq:wkF}) for any function~$g(\theta)$.
When $F_k = c_1 g(\theta) k^{-7/2}$, the energy flux in FFF
interactions per unit mass per unit solid angle in $k$-space,
$\epsilon$, can be obtained in the same way as for weak acoustic
turbulence~\cite{zak70}, and is given by $\epsilon = 9\pi^2 c_1^2
\sin^2 \theta g^2 c_2/16 v_A$, where $c_2 = \int_0^\infty dx
\ln(1+x)[x(1+x)]^{-5/2}[(1+x)^{9/2}-x^{9/2}-1] \simeq 26.2$.  If the
fast waves were forced isotropically and non-FFF interactions were
ignored, then~$g= 1/\sin\theta$ in steady state so that~$\epsilon$ is
independent of~$\theta$. The time scale~$\tau_F$ for FFF interactions
to transfer fast-wave energy from~$k$ to $2k$ can be estimated by
dividing the fast-wave energy per unit solid angle between~$k$
and~$2k$ by the energy flux~$\epsilon$. Ignoring numerical
coefficients, one obtains $\tau_F \sim v_A/[c_1 \sin^2\theta g(\theta)
  k^{1/2}]$ for $F_k = c_1 g(\theta) k^{-7/2}$.  If the fast-wave
energy were dominated by wavenumbers of order some characteristic wave
number~$k_0$, with $F_k = c_1g(\theta) k^{-7/2}$ for $k\gtrsim k_0$,
then $c_1 \sim (\delta v_{{\rm rms},F})^2 k_0^{1/2}$, where $(\delta
v_{{\rm rms},F})^2$ is the energy per unit mass in fast waves.  In
this case, $\tau_F \sim v_A/[(\delta v_{{\rm rms},F})^2 (k_0 k)^{1/2}
  \sin^2\theta g(\theta)]$ for~$k\gg k_0$.

The terms in equations~(\ref{eq:wkA}) and (\ref{eq:wkF}) containing
$M_1$ through~$M_{17}$ correspond to three-wave interactions involving
both Alfv\'en waves and fast waves (``AAF and AFF interactions'').
Such interactions exchange energy between fast waves and Alfv\'en
waves within resonant wavenumber triads. At small~$\theta$, the
frequencies of fast waves and Alfv\'en waves are comparable, and AAF
and AFF interactions are efficient.  For example, if $F_k=c_1 k^{-7/2}
\sin^{-1}\theta$ and $A_k\ll F_k$ at small~$\theta$, then when
$\theta\ll 1$ the largest contribution to~$\partial F_k/\partial t$
comes from the term proportional to~$M_{13} F_q F_k$ and is~$
-F_k/\tau_{AF}$, where $\tau_{AF} = (15 v_A \sin\theta)/(23 \pi^2 c_1
k^{1/2}) $ to lowest order in~$\theta$, a time scale that is~$\ll
\tau_F$.  The energy lost by fast waves in this case is transferred
primarily to Alfv\'en waves at the same wavenumber through the term in
equation~(\ref{eq:wkA}) containing $M_8F_p F_q$. If $A_k$ grows until
$A_k=F_k$ at small~$\theta$, then the term containing $M_{13}F_qF_k$
is cancelled by the term containing $M_{13} A_p F_k$ to lowest order
in~$\theta$, largely stemming the loss of fast-wave energy. AAF and
AFF interactions thus act to make $A_k \simeq F_k$ at small~$\theta$.
However, the constant-energy-flux solution $A_k \simeq F_k \propto
k^{-7/2}$ is unsustainable, because as $k$ increases energy is lost
from the small-$\theta$ part of $k$-space to high~$k_\perp$ through
AAA interactions faster than it is replenished from small~$k$ by FFF
interactions ($\tau_F\propto k^{-1/2}$,~$\tau_A \propto k^{-1}$).  The
energy flux in FFF interactions at small~$\theta$ must thus decrease
with~$k$ as fast-wave energy is drained into Alfv\'en waves and then
transferred out to large~$k_\perp$. This process causes $F_k$ to
steepen relative to~$k^{-7/2}$ at small~$\theta$, and results in
Alfv\'en-wave energy at $|k_z|\gg k_0$.  On the other hand,
for~$\theta \gtrsim 45^\circ$, the frequencies of Alfv\'en waves and
fast waves differ considerably, and AAF and AFF interactions are
unable to make~$A_k\simeq F_k$ at~$k \gg k_0$.  In this part of
$k$-space, AAA and FFF interactions dominate the right-hand sides of
equations~(\ref{eq:wkA}) and (\ref{eq:wkF}), so that $F_k \propto
k^{-7/2}$ and $A_k = h(k_z) k_\perp^{-3}$ within the inertial range,
where~$h(k_z)$ is some (decreasing) function of~$|k_z|$.

\begin{figure}[h]
\includegraphics[width=3.1in]{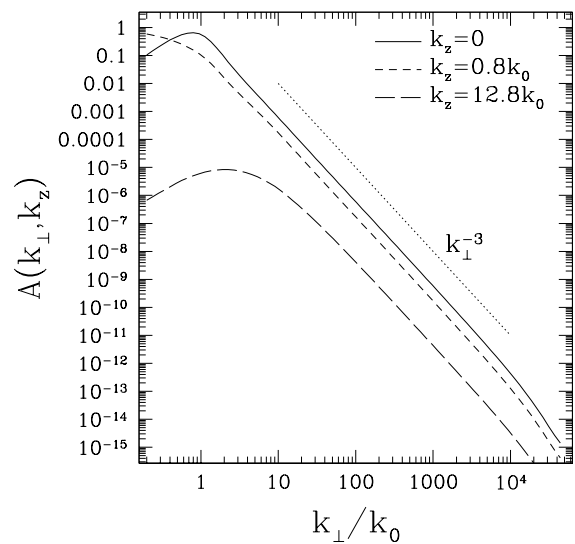}
\caption{Alfv\'en-wave power spectrum as a function of~$k_\perp$
at different~$k_z$.
\label{fig:f2} }
\end{figure}

\begin{figure}[h]
\includegraphics[width=3.1in]{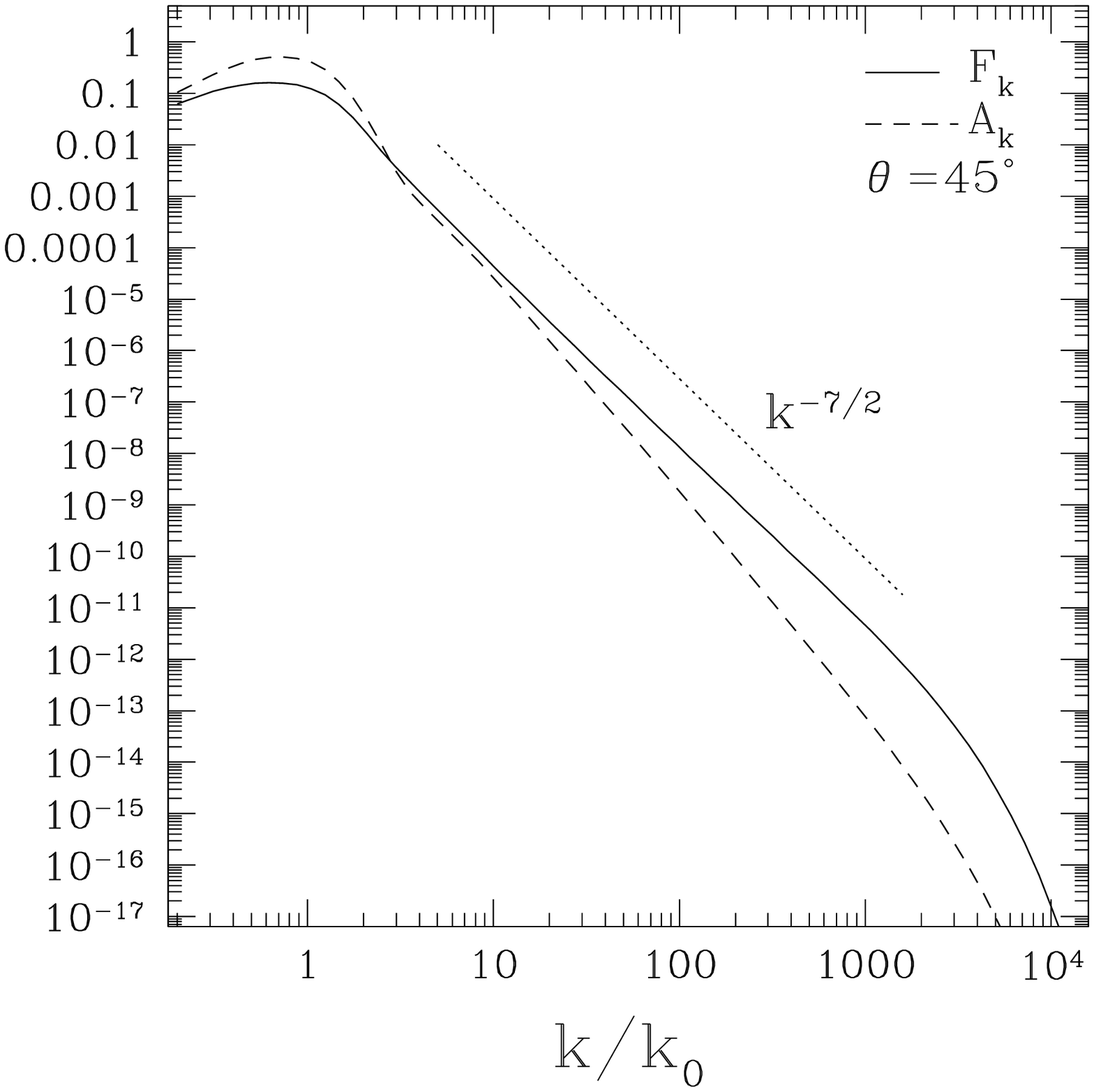}
\includegraphics[width=3.1in]{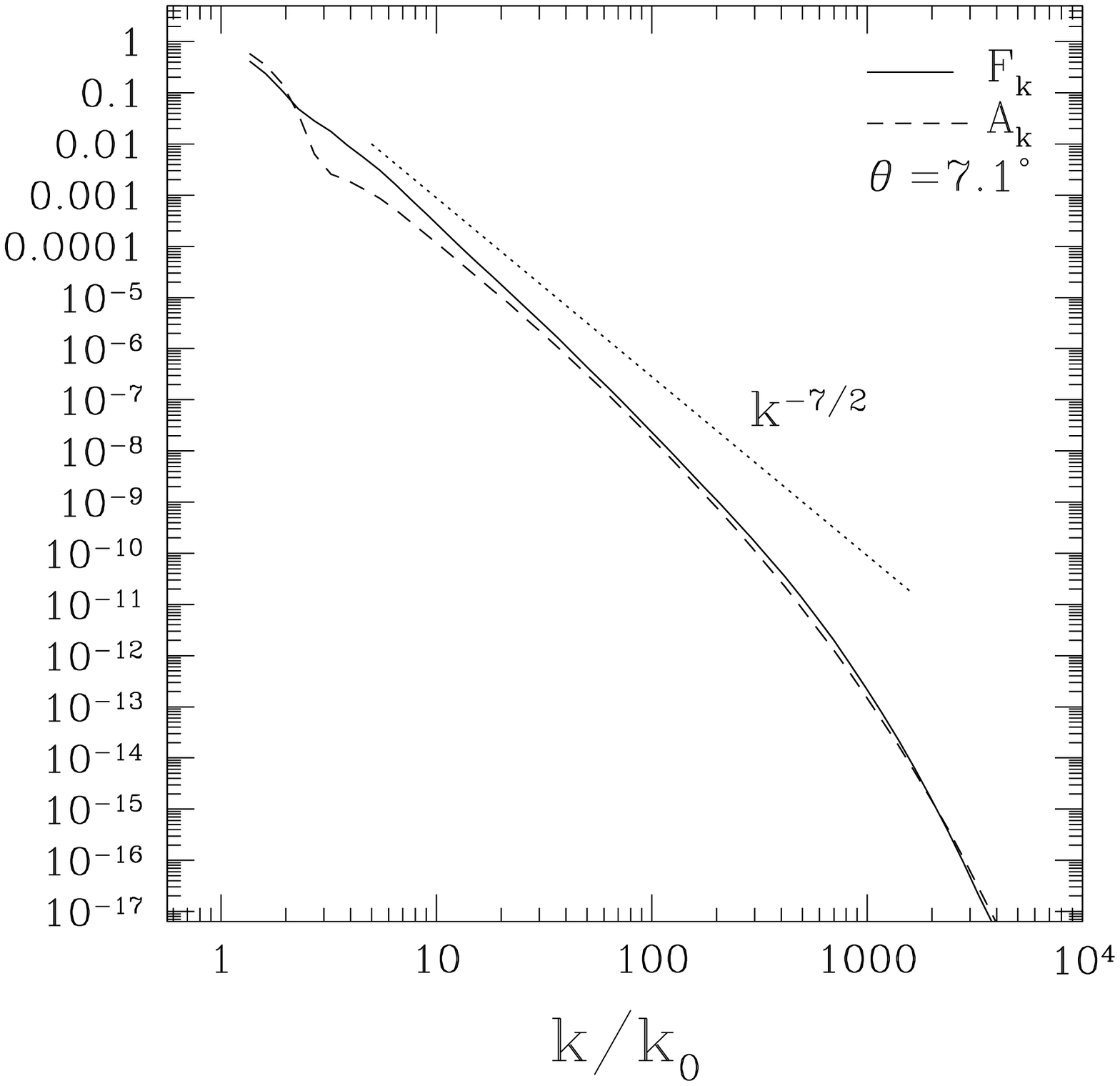}
\caption{{\em Top panel:} power spectra as a function of~$k$ at~$\theta = 45^\circ$.
{\em Bottom panel:} power spectra as a function of~$k$ at~$\theta = 7.1^\circ$.
\label{fig:f3} }
\end{figure}

To obtain quantitative solutions for~$A_k$ and $F_k$,
equations~(\ref{eq:wkA}) and (\ref{eq:wkF}) are integrated forward in
time numerically with initial spectra $A_k = F_k = k^2
\exp(-k^2/k_0^2)$. The isotropic forcing term $c_3 k^2
\exp(-k^2/k_0^2)$ is added to the right-hand sides of both
equations~(\ref{eq:wkA}) and (\ref{eq:wkF}). The dissipation terms~$-
c_4 k^2 A_k$ and $-c_4 k^2 F_k$ are added to the right-hand sides of
equations~(\ref{eq:wkA}) and (\ref{eq:wkF}) respectively, with~$c_4$
chosen so that in steady state dissipation truncates the spectra at a
wavenumber that is~$\gg k_0$. The numerical method
conserves energy to machine accuracy in the absence of dissipation and
forcing and will be described in a future publication. Steady-state
spectra at late times are plotted in figures~\ref{fig:f2}
and~\ref{fig:f3} and are consistent with the qualitative picture
described above. 
The Alfv\'en-wave spectra are $\propto h(k_z) k_\perp^{-3}$ for
$k_\perp \gg |k_z|$ within the inertial range.  Although $h(k_z)$
decreases with increasing~$|k_z|$, there is some Alfv\'en-wave energy at
$|k_z|\gg k_0$, in contrast to the case of weak incompressible MHD
turbulence.~\cite{gal00} At $\theta = 45^\circ$, $F_k$ is~$\propto
k^{-7/2}$ and~$A_k$ drops off more steeply
than~$k^{-7/2}$. At~$\theta=7.1^\circ$, $F_k$ falls off more rapidly
than~$k^{-7/2}$ and AAF and AFF interactions keep~$A_k \simeq F_k$.
The cascade of fast-wave energy to high frequencies was found
previously by~\cite{cho03} in direct numerical simulations with
strongly turbulent Alfv\'en waves and slow waves. In contrast to this
Letter, these authors found an isotropic $k^{-7/2}$ fast-wave spectrum.

The phenomenology described above can be applied more generally.  For
example, if the $z$-component of the phase velocity, $v_{{\rm ph},z}$,
were initially positive for all the excited waves, and if there were
no mechanism for generating Alfv\'en-waves with $k_z=0$ and $v_{\rm
  ph,z}<0$, then there would be no AAA interactions. In this case, FFF
interactions would still transfer fast-wave energy to high
frequencies, and AAF and AFF interactions would still cause $A_k$ and
$F_k$ to become approximately equal at small~$\theta$, but the
Alfv\'en-wave energy would not be swept out to large~$k_\perp$ by AAA
interactions. For waves with~$v_{{\rm ph},z}>0$, one would thus expect
$F_k$ to obtain a constant-energy-flux $k^{-7/2}$ scaling for
all~$\theta$ with $A_k \simeq F_k$ at small~$\theta$. As a second
example, if the initial excitation were primarily in Alfv\'en waves,
as may be the case in the corona~\cite{hol02}, and if $A_k$ were
quasi-isotropic at~$k\sim k_0$, then AAF and AFF interactions would
generate significant fast-wave energy at~$k\sim k_0$, and FFF
interactions would subsequently transfer fast-wave energy to higher
frequencies.  As a final example, if $\delta v_{\rm rms} \ll v_A$ but
the Alfv\'en waves at small~$|k_z|$ became strongly turbulent at
$k_\perp$ larger than some transition wave number $k_{\rm tr}$, as
in~\cite{gol97}, then collisions between oppositely directed Alfv\'en
wave packets would still transfer Alfv\'en wave energy at any~$k_z$ to
larger $k_\perp$, but the cascade time for this process at $k_\perp >
k_{\rm tr}$ would change from $\tau_A$ to a new value, $\tau_{A,{\rm
    str}} \propto k_\perp^{-2/3}$~\cite{gol97}. The Alfv\'en waves in
most of $k$-space and the fast waves would still be weakly turbulent
because the linear periods of these waves would still be much shorter
than the nonlinear time scales, and much of the weak-turbulence picture
would still apply. In particular, $F_k$ would be~$\propto k^{-7/2}$
for $\theta\gtrsim 45^\circ$, and $A_k$ and $F_k$ would be steeper
than~$k^{-7/2}$ at small~$\theta$ and large~$k$ since $\tau_A$,
$\tau_{A,{\rm str}}$, and~$\tau_{AF}$ are~$\ll \tau_F$ in that part
of~$k$-space.

I thank Sebastian Galtier, Ellen Zweibel, Amitava Bhattacharjee, Marty
Lee, and Joe Hollweg for helpful discussions.  This work was supported
by NASA under grant NNG 05GH39G and by NSF under grant AST
05-49577.

\end{document}